\documentclass{iau}
\usepackage{graphicx}
\usepackage{slashbox}
\title[Impact flux of asteroids]
{Impact flux of asteroids and water transport to the habitable zone in binary star systems}

\author[D. Bancelin \etal]   
{D. Bancelin$^{1,3}$, E. Pilat-Lohinger$^{2,1}$, S. Eggl$^3$ \and R. Dvorak$^1$}

\affiliation{$^1$Institute of Astrophysics, University of Vienna, Austria \\
email: {\tt david.bancelin@univie.ac.at } \\[\affilskip]
$^2$Institute of Physics, University of Graz, Austria \\[\affilskip]
$^3$IMCCE, Paris Observatory, France }

\pubyear{2014}
\volume{310}  
\pagerange{xxx--xxx}
\jname{Complex Planetary Systems}
\editors{Z. Knezevic \& A. Lema\^itre}

\begin{document}

\maketitle

\begin{abstract}
By now, observations of exoplanets have found more than 50 binary star systems hosting 71 planets. We expect these
numbers to increase as more than 70\% of the main sequence stars in the solar neighborhood are members of
binary or multiple systems. The planetary motion in such systems depends strongly on both the parameters of the
stellar
system (stellar separation and eccentricity) and the architecture of the planetary system (number of planets and their
orbital behaviour). In case a terrestrial planet moves in the so-called habitable zone (HZ) of its host star, the
habitability of this planet depends on many parameters. A crucial factor is certainly the amount of water. We
investigate in this work the transport of water from beyond the snow-line to the HZ in a binary star system and compare
it to a single star system.

\end{abstract}

\firstsection 
\section*{Introduction}
Water is the main ingredient defining a habitable planet. Therefore, the main question we would like to answer in our
study is if a dry or almost dry planet can be fed with water by a bombardement of wet small bodies in binary
systems.
First simulations of planetary formation in such systems show the stochastic behaviour of the water to mass ratio of
planetary embryos (\cite{haghighipour07})
. After the planetary formation, a remnent disc of
small bodies
can be found around the main star. It mainly contains asteroids and comets whose initial water distribution depends on
their relative position to the snow-line. In our study, we mainly tackle the question of the amount of water available
in the HZ. To this purpose, we
consider various binary star systems where both stars are G types. We analyse the dynamics of a ring
containing 100
asteroids (with maximum mass equal to the mass of Ceres) initially placed beyond the snow-line (2.7 au for a primary G
star)
moving under the gravitational perturbation of the binary stars and a Jupiter at 5.2 au. The systems were studied
numerically for 10 Myr. To make this study statistical, each disk is cloned 100 times. Thus the statistics is made
with 10000 asteroids.
\section*{Results}
\begin{figure}
\centering{
    \begin{tabular}{cc}
      \includegraphics[angle=-90, width=0.5\textwidth]{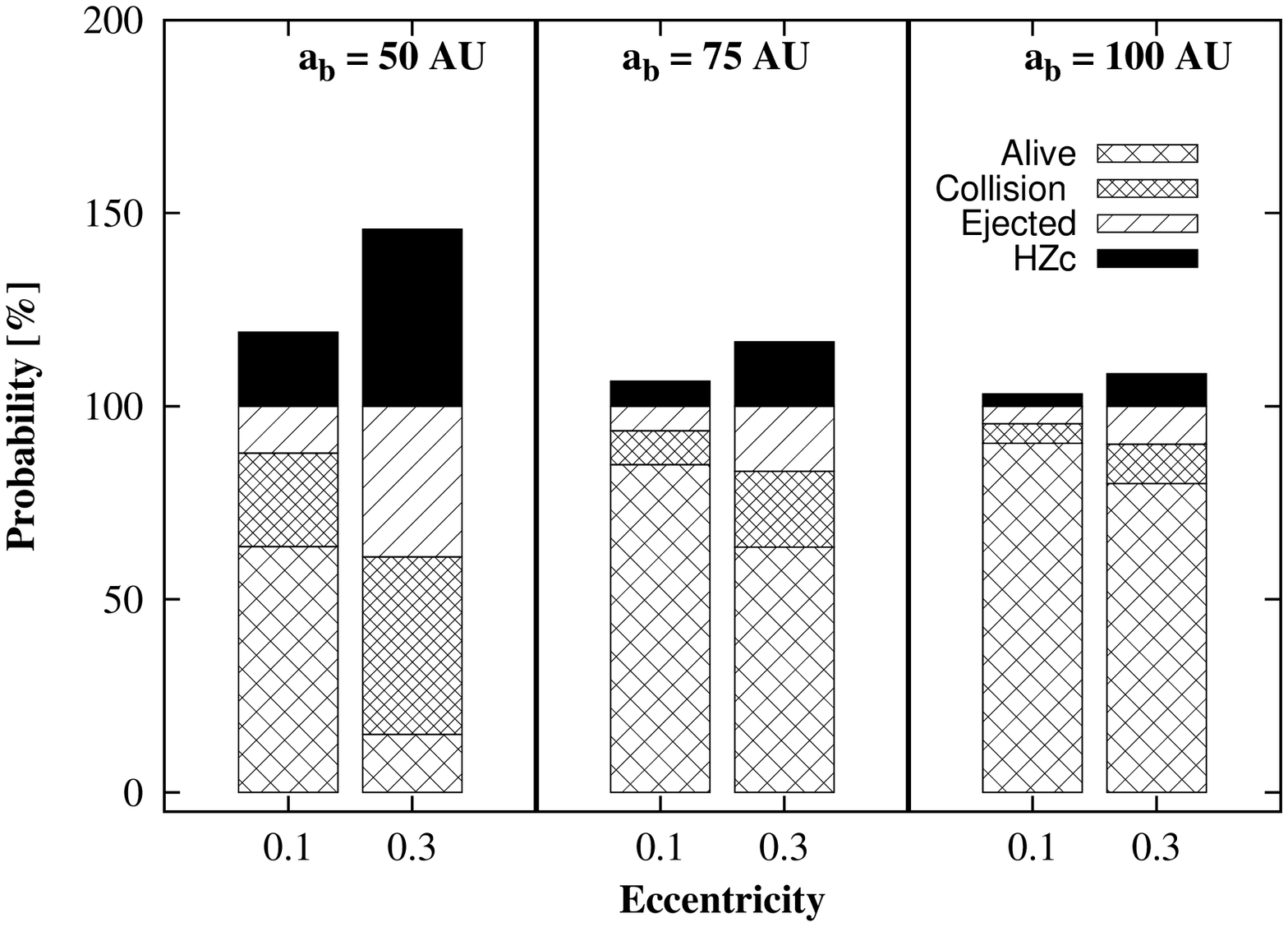} &
      \includegraphics[angle=-90, width=0.5\textwidth]{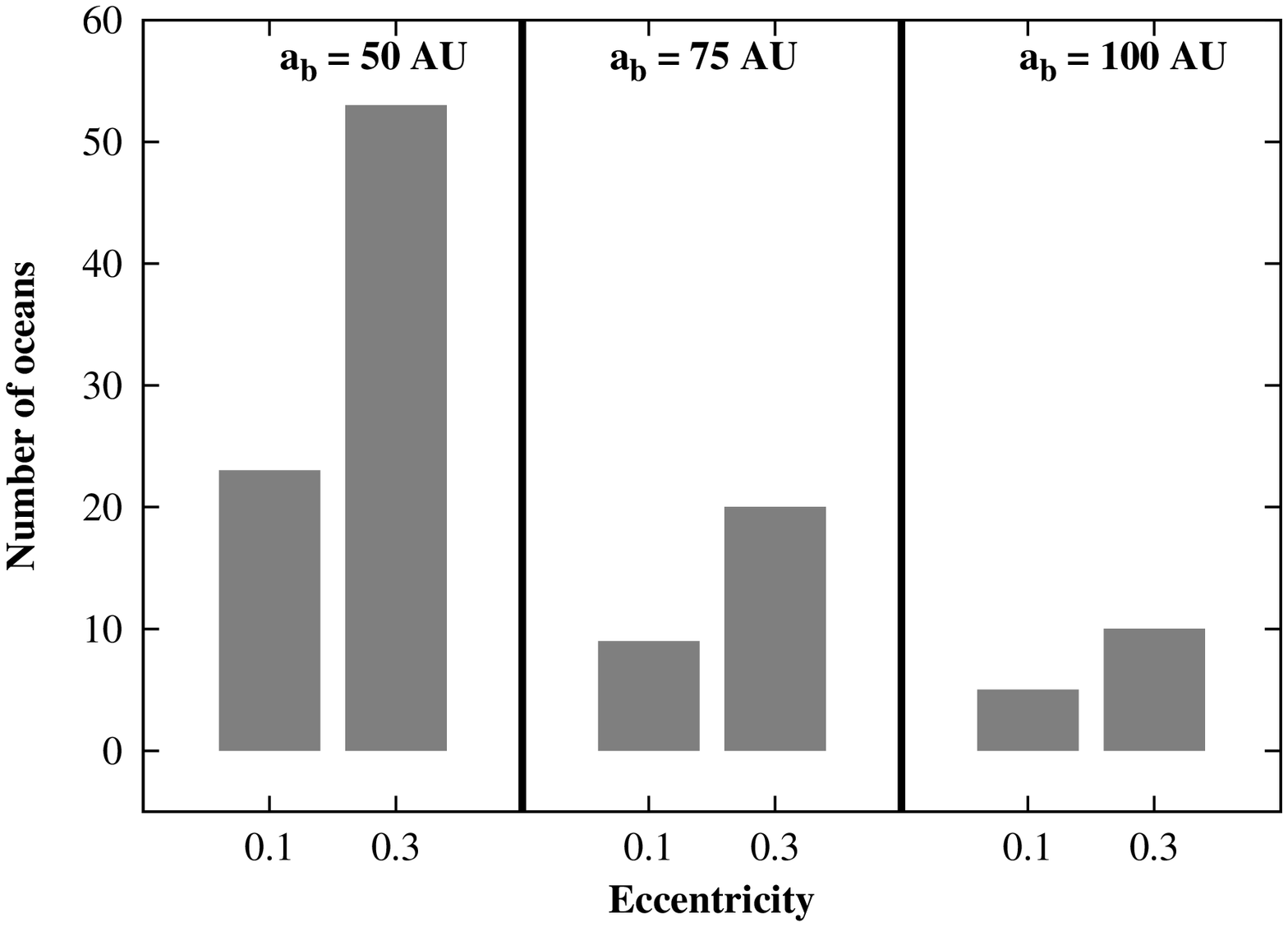} \\
    \end{tabular}}
    \caption{Left panel: Statistics on the asteroid's dynamics as a function of the binary's
eccentricity and semi-major axis a$_{\scriptscriptstyle b}$. Each pattern refers to the probability for an asteroid
initially in the ring to collide with the stars or Jupiter, be ejected or still be present in the system after 10 Myr.
HZc refers to the probability for an asteroid to cross the HZ. Right panel: Quantity of water (in ocean units)
transported to the HZ assuming also a water loss due to ice sublimation.} \label{F:impactor}
\end{figure}

\noindent
 Figure \ref{F:impactor} (left panel) shows the statistics on the asteroids dynamics. The perturbations of both Jupiter
and the binary companion will drastically influence the asteroids in the ring and may increase their 
eccentricities and evolve beyond their
initial semi-major axis borders. This behaviour is highlighted by the 
increasing value of the probability for an asteroid to cross the HZ as a function of the periapsis distance. They are
called
habitable zone crossers (HZc). As a consequence, the ring will be depopulated because the dynamics will induce
ejections 
of asteroids and collisions with Jupiter
and the stars. Therefore, the probability for an asteroid initially in the ring 
to stay in the system ("alive") after 10 Myr will
decrease with the periapsis distance of the secondary. Assuming an initial water content of 10\% for each asteroid, we
derived the water to mass fraction when it first
enters the HZ. This corresponds to the maximum value of transported water by this asteroid. We also assume a water loss
process due to ice sublimation caused by the radiation of both stars. Indeed, we estimated up to 14\% contribution
of the secondary star during the sublimation process. 
The right panel of Fig. \ref{F:impactor} shows the total amount of
water transported (in ocean units) by the HZc among the 10000 asteroids of all rings. As we can see, a binary companion
on an
eccentric orbit helps for a more efficient transport. Indeed, the ring is highly perturbed because the presence of
Jupiter will induce secular perturbations. Only a few oceans can be transported for the most distant secondary
stars because of the small flux of incoming asteroids in the HZ. This means that the timescale to drastically increase
the transport of water must be beyond 10 Myr. This test shows how efficient a system can be to rapidly
transport water to the HZ. However, we estimated that the maximum duration for the last HZc to bring water to
the HZ is less than 0.25 Myr. This suggest a transport in more than one step as some systems  still have more than 50\%
of asteroids in the ring (i.e. they have never crossed the HZ). Finally, we estimate that the water transport is
$\sim$4--­5 less efficient if the
system is not a binary.

\noindent

\section*{Acknowledgments}
\noindent
DB and EPL acknowledge the support of the 
FWF project S11608-N16 and S11603-N16 (NFN-subprojects) and
the Vienna Scientific Cluster (VSC) for the computational ressources.

\end{document}